\documentclass{article}
\begin{document}

\title{Microscopic Calculation of the Dielectric Susceptibility 
Tensor for Coulomb Fluids II}

\author{
B. Jancovici$^1$ and L. {\v S}amaj$^{1,2}$
}

\maketitle

\begin{abstract}
For a Coulomb system contained in a domain $\Lambda$, the
dielectric susceptibility tensor $\chi_{\Lambda}$ is defined
as relating the average polarization in the system to a constant
applied electric field, in the linear limit.
According to the phenomenological laws of macroscopic electrostatics,
$\chi_{\Lambda}$ depends on the specific shape of the domain
$\Lambda$.
In this paper we derive, using the methods of equilibrium 
statistical mechanics in both canonical and grand-canonical
ensembles, the shape dependence of $\chi_{\Lambda}$ and
the corresponding finite-size corrections to the thermodynamic
limit, for a class of general $\nu$-dimensional $(\nu \ge 2)$
Coulomb systems, of ellipsoidal shape, being in the conducting state.
The microscopic derivation is based on a general principle:
the total force acting on a system in thermal equilibrium is zero.
The results are checked in the Debye-H\"uckel limit. 
The paper is a generalization of a previous one
[L. \v Samaj, {\it J. Stat. Phys.} {\bf 100}:949 (2000)], dealing with
the special case of a one-component plasma in two dimensions.
In that case, the validity of the presented formalism has already
been verified at the exactly solvable (dimensionless) 
coupling $\Gamma=2$.

\end{abstract}

\vfill

\noindent $^1$Laboratoire de Physique Th\'eorique, Universit\'e
Paris-Sud, B\^atiment 210, 91405 Orsay Cedex, France
(Unit\'e Mixte de Recherche no. 8627 - CNRS); 
e-mail: Bernard.Jancovici@th.u-psud.fr

\noindent $^2$Institute of Physics, Slovak Academy of Sciences,
D\'ubravsk\'a cesta 9, 84511 Bratislava, Slovakia;
e-mail: fyzimaes@savba.sk

\newpage

\renewcommand{\theequation}{1.\arabic{equation}}
\setcounter{equation}{0}

\section{INTRODUCTION}
For systems with short-ranged pair interactions among 
constituents defined in a specifically shaped domain $\Lambda$,
the thermodynamic limit of an intensive quantity does not depend
in general on the shape of the domain $\Lambda$ and
on the conditions at the boundary $\partial \Lambda$ given by
the surrounding medium. 
This is no longer true in the case of macroscopic systems with 
long-ranged pair interactions.
A typical example is the domain-shape dependence of the 
dielectric susceptibility tensor for conductors
predicted by the phenomenological laws of electrostatics
\cite{Landau}.
The aim of this paper is to derive rigorously and precisely, 
using the methods of equilibrium statistical mechanics in 
both canonical and grand-canonical ensembles, the shape dependence 
of the dielectric susceptibility and the corresponding finite-size 
corrections to the thermodynamic limit, for a class of general 
classical $\nu$-dimensional microscopic Coulomb systems being 
in the conducting state.
The case $\nu=1$ has special features and 
will not be discussed here.
The paper is a generalization of the previous one \cite{Samaj1},
referred to as I, which was devoted to the microscopic derivation
of the dielectric susceptibility for the special case of a 
one-component plasma in two dimensions (2D).
 
In dimension $\nu$, the Coulomb potential $v$ at a spatial position
${\bf r} = (r^1,r^2,\ldots,r^{\nu})$, induced by a unit charge
at the origin ${\bf 0}$, is the solution of the Poisson equation
\begin{equation} \label{1.1}
\Delta v({\bf r}) = - s_{\nu} \delta({\bf r})
\end{equation} 
where $s_{\nu}=2\pi^{\nu/2}/\Gamma(\nu/2)$ is the surface area 
of the $\nu$-dimensional unit sphere.
Explicitly,
\begin{equation} \label{1.2}
v({\bf r}) = \left\{
\begin{array}{cl}
-\ln(r/r_0) & \mbox{if $\nu=2$,} \\ & \\
\displaystyle{\frac{r^{2-\nu}}{\nu-2}} & \mbox{otherwise}
\end{array} 
\right.
\end{equation}
Here, $r =\vert {\bf r} \vert$ and $r_0$ is an arbitrary length scale.
The corresponding force ${\bf F}({\bf r}) = - \nabla v({\bf r})$
reads
\begin{equation} \label{1.3}
{\bf F}({\bf r}) = \frac{{\bf r}}{r^{\nu}}
\end{equation}
In a $\nu$-dimensional space, the definition of the
Coulomb potential (\ref{1.1}) implies in the Fourier space the 
characteristic small-$k$ behavior ${\hat v}({\bf k}) \propto 1/k^2$.
This maintains many generic properties (like screening) 
of ``real'' 3D Coulomb systems.

We consider general Coulomb systems consisting of $M$ mobile 
species $\alpha = 1,\ldots,M$ 
with the corresponding charges $q_{\alpha}$,
embedded in a fixed uniform background of charge density $\rho_b$.
The most studied models are the one-component plasma (OCP) and
the symmetric two-component plasma (TCP).
The OCP corresponds to $M=1$ with $q_1=q$ and $\rho_b$ of 
opposite sign; it may be convenient to define a ``background density''
$n_b$ by $\rho_b = -q n_b$.
The symmetric TCP corresponds to $M=2$ with $q_1=q, q_2=-q$ and
$\rho_b=0$.
The system is contained in a domain $\Lambda$ of specified shape
with a smooth boundary $\partial \Lambda$.
The surrounding medium is for simplicity a vacuum producing
no image forces.
The fixed background produces the one-particle potential
$\rho_b \phi_b({\bf r})$ where
\begin{equation} \label{1.4}
\phi_b({\bf r}) = \int_{\Lambda} {\rm d}^{\nu} r'
v\left(\vert {\bf r}-{\bf r}'\vert \right)
\end{equation}
The corresponding electric field is $\rho_b {\bf E}_b({\bf r})$ where
\begin{equation} \label{1.5}
{\bf E}_b({\bf r}) = - \nabla \phi_b({\bf r})
= \int_{\Lambda} {\rm d}^{\nu} r' 
\frac{{\bf r}-{\bf r}'}{\vert {\bf r}-{\bf r}'\vert^{\nu}}
\end{equation}
The energy of a configuration $\{ {\bf r}_i, q_{\alpha_i} \}$
of the charged particles plus the background is
\begin{equation} \label{1.6}
E = \sum_{i<j} q_{\alpha_i} q_{\alpha_j} v(\vert {\bf r}_i-{\bf r}_j\vert) 
+ \rho_b \sum_i q_{\alpha_i} \phi_b({\bf r}_i) + E_{b-b}
\end{equation}
Since the backgroud-background interaction energy term $E_{b-b}$ 
does not depend on the particle coordinates, its particular value is 
irrelevant in the calculation of particle distribution functions.
In the case of point particles, for many-component systems with at least
two oppositely charged species,  
the singularity of $v(r)$ (\ref{1.2}) at the origin prevents
the thermodynamic stability against the collapse of positive-negative
pairs of charges: in two dimensions for small enough temperatures,
in three and higher dimensions for any temperature.
However, in those cases, one can introduce short-range repulsive
interactions which prevent the collapse. The derivations which follow
allow for such interactions.

The Coulomb system in the domain $\Lambda$ at inverse temperature 
$\beta$ will be considered in both canonical (fixed particle
numbers) and grand canonical (fixed species chemical potentials) 
ensembles.
The thermal average will be denoted by $\langle \cdots \rangle$.
In terms of the microscopic density of particles of species $\alpha$,
${\hat n}_{\alpha}({\bf r}) = \sum_i \delta_{\alpha,\alpha_i}
\delta({\bf r}-{\bf r}_i)$, the microscopic densities of the total
particle number and charge are defined respectively by
\begin{equation} \label{1.7}
{\hat n}({\bf r}) = \sum_{\alpha} {\hat n}_{\alpha}({\bf r}),
\quad \quad {\hat \rho}({\bf r}) = \sum_{\alpha} q_{\alpha}
{\hat n}_{\alpha}({\bf r})
\end{equation}
At one-particle level, the total particle number and charge
densities are given respectively by
\begin{equation} \label{1.8}
n({\bf r}) = \langle {\hat n}({\bf r}) \rangle , \quad \quad
\rho({\bf r}) = \langle {\hat\rho}({\bf r}) \rangle
\end{equation}
At two-particle level, one introduces the two-body densities
\begin{eqnarray} 
n_{\alpha\alpha'}^{(2)}({\bf r},{\bf r}') & = &
\left\langle \sum_{i\ne j} \delta_{\alpha,\alpha_i} 
\delta_{\alpha',\alpha_j} \delta({\bf r}-{\bf r}_i)
\delta({\bf r}'-{\bf r}_j) \right\rangle \nonumber \\
& = & \langle {\hat n}_{\alpha}({\bf r}) {\hat n}_{\alpha'}({\bf r}')
\rangle - \langle {\hat n}_{\alpha}({\bf r}) \rangle 
\delta_{\alpha,\alpha'} \delta({\bf r}-{\bf r}') \label{1.9}
\end{eqnarray}
The corresponding Ursell functions are defined by
\begin{equation} \label{1.10}
U_{\alpha\alpha'}({\bf r},{\bf r}') = 
n_{\alpha\alpha'}^{(2)}({\bf r},{\bf r}') - n_{\alpha}({\bf r})
n_{\alpha'}({\bf r}') 
\end{equation}
and the truncated charge-charge structure function by
\begin{eqnarray} 
S({\bf r},{\bf r}') & = &
\langle {\hat\rho}({\bf r}) {\hat\rho}({\bf r}')\rangle^{\rm T}
\nonumber \\ & \equiv & 
\langle {\hat\rho}({\bf r}) {\hat\rho}({\bf r}')\rangle - 
\langle {\hat\rho}({\bf r}) \rangle \langle {\hat\rho}({\bf r}') \rangle
\label{1.11}
\end{eqnarray}

The small-$k$ behavior of the Fourier transform of 
the Coulomb potential gives rise to exact moment constraints 
for the charge structure function $S$ (see review \cite{Martin}).
In the bulk, $\lim_{\Lambda\to {\bf R}^{\nu}} 
S_{\Lambda}({\bf r},{\bf r}') = S(\vert {\bf r}-{\bf r}'\vert)$
obeys the Stillinger-Lovett screening rules \cite{Stillinger1,Stillinger2}
which imply the zeroth-moment (electroneutrality) condition
\begin{equation} \label{1.12}
\int {\rm d}^{\nu} r S({\bf r}) = 0
\end{equation}
and the second-moment condition
\begin{equation} \label{1.13}
\beta \int {\rm d}^{\nu} r \vert {\bf r} \vert^2 S({\bf r}) = 
- \frac{2\nu}{s_{\nu}}
\end{equation}

For finite systems, the analog of the zeroth-moment sum rule
\begin{equation} \label{1.14}
\int_{\Lambda} {\rm d}^{\nu}r S({\bf r},{\bf r}')
= \int_{\Lambda} {\rm d}^{\nu}r' S({\bf r},{\bf r}') = 0
\end{equation} 
holds only in the canonical ensemble where it reflects
the trivial fact that the total charge in the domain $\Lambda$
is fixed.
In the grand canonical ensemble, the system is expected
to exhibit charge fluctuations \cite{Lieb}, in which case (\ref{1.14})
does not hold.
The information analogous to the bulk second-moment condition 
(\ref{1.13}) is contained in the dielectric susceptibility
tensor $\chi_{\Lambda}$. 
Let us use the notation
\begin{equation} \label{1.15}
{\hat P}^i = \int_{\Lambda} {\rm d}^{\nu}r\, r^i {\hat\rho}({\bf r})
\quad \quad i = 1,\ldots,\nu
\end{equation}
for the $i$th component of the total polarization in the system.
The tensor $\chi_{\Lambda}$ is defined as relating the average 
polarization to a constant applied field ${\bf E}_0$,
in the linear limit:
\begin{equation} \label{1.16}
\frac{\langle {\hat P}^i \rangle_{{\bf E}_0}}{\vert \Lambda\vert} =
\sum_{j=1}^{\nu} \chi_{\Lambda}^{ij} E^j_0
\end{equation} 
The linear response theory expresses the $\chi_{\Lambda}$-components as
\begin{eqnarray} 
\chi_{\Lambda}^{ij} & = & \frac{\beta}{\vert\Lambda\vert}
\left( \langle {\hat P}^i {\hat P}^j \rangle - 
\langle {\hat P}^i \rangle \langle {\hat P}^j \rangle \right) 
\nonumber \\ & = & \frac{\beta}{\vert\Lambda\vert} 
\int_{\Lambda} {\rm d}^{\nu}r_1  \int_{\Lambda} {\rm d}^{\nu}r_2
\, r_1^i r_2^j S({\bf r}_1,{\bf r}_2) \label{1.17}
\end{eqnarray}
where $\langle \cdots \rangle$ is an average defined for ${\bf E}_0=0$.
In the canonical ensemble where the sum rule (\ref{1.14}) applies,
the tensor components $\chi_{\Lambda}^{ij}$ are expressible in another
equivalent way
\begin{equation} \label{1.18}
\chi_{\Lambda}^{ij} = - \frac{\beta}{2\vert \Lambda\vert}
\int_{\Lambda} {\rm d}^{\nu}r_1  \int_{\Lambda} {\rm d}^{\nu}r_2
(r_1^i- r_2^j)^2 S({\bf r}_1,{\bf r}_2)
\end{equation}
As $\Lambda \to {\bf R}^{\nu}$ one might naively expect that
only the diagonal components 
$\chi^i = \lim_{\Lambda\to {\bf R}^{\nu}} \chi_{\Lambda}^{ii}$
$(i=1,\ldots,\nu)$ survive and, according to the bulk second-moment
sum rule (\ref{1.13}), that they tend to 
the uniform ``Stillinger-Lovett'' (SL) value
\begin{equation} \label{1.19}
\chi_{SL}^i = - \frac{\beta}{2} \int {\rm d}^{\nu}r 
\left( r^i \right)^2 S({\bf r}) = \frac{1}{s_{\nu}}
\end{equation}
which does not depend on the shape of $\Lambda$. 
This is indeed true for a boundary-free domain like the surface 
of a sphere.
As is explained below, relation (\ref{1.19}) no longer holds
in a geometry with a boundary.
 
According to phenomenological electrostatics, based on plausible
but not rigorously justified arguments, the dielectric susceptibility
$\chi$ of a macroscopic system is related to its dielectric constant
$\epsilon$.
For the considered Coulomb plasma in a conducting state,
the equality $\epsilon^{-1}=0$ implies
\begin{equation} \label{1.20}
\chi_{\Lambda}^{ij} = \frac{1}{s_{\nu}}
\left( T_{\Lambda}^{-1} \right)^{ij}
\end{equation}
where $T_{\Lambda}$ is the size-invariant but shape-dependent
depolarization tensor with position-independent components
\begin{equation} \label{1.21}
T_{\Lambda}^{ij} = - \frac{1}{s_{\nu}} \int_{\Lambda} {\rm d}^{\nu}r
\frac{\partial^2 v({\bf r})}{\partial r^i \partial r^j}
\end{equation}
Without any loss of generality one can choose a coordinate system
in which $T_{\Lambda}$ is diagonal, 
$T_{\Lambda}^{ij}=T_{\Lambda}^i\delta_{ij}$, and consequently
$\chi_{\Lambda}$ is also diagonal, 
$\chi_{\Lambda}^{ij}=\chi_{\Lambda}^i\delta_{ij}$.
Then, Eq. (\ref{1.20}) takes the form
\begin{equation} \label{1.22}
\chi_{\Lambda}^i = \frac{1}{s_{\nu} T_{\Lambda}^i}
\end{equation}
For $\nu$-dimensional ellipsoidal domains
\cite{Landau} which will be of interest in this work,
the $T_{\Lambda}$-components (\ref{1.21}) are expressible
in an alternative form as
\begin{equation} \label{1.23}
T_{\Lambda}^{ij} = - \frac{1}{s_{\nu}} 
\frac{\partial^2}{\partial r^i \partial r^j} \phi_b({\bf r})
\end{equation} 
with $\phi_b({\bf r})$ defined by (\ref{1.4}),
where ${\bf r}$ is an arbitrary point in $\Lambda$.
With regard to the Poisson equation (\ref{1.1}), the diagonal
elements of $T_{\Lambda}$ are constrained by
$\sum_{i=1}^{\nu} T_{\Lambda}^i = 1$.
In the special isotropic case of $\nu$-dimensional spheres,
$T_{\Lambda}^i=1/\nu$ and the consequent
$\chi_{\Lambda}^i = \nu/s_{\nu}$ is $\nu$ times $\chi_{SL}^i$
of Eq. (\ref{1.19}).

The discrepancy between the naive prediction of statistical mechanics
(\ref{1.19}) and phenomenological electrostatics (\ref{1.22})
was explained in a nice series of papers 
\cite{Choquard1}-\cite{Choquard3} by Choquard et al.
The point is that the susceptibility is made up of a bulk 
contribution, which saturates quickly to the SL value (\ref{1.19}),
and of a surface contribution.
The surface contribution does not vanish in the thermodynamic limit
due to the inverse-power-law behavior of the charge structure 
function at large distances along the boundary.
Summing up both contributions one gets instead of (\ref{1.19})
the shape-dependent result of macroscopic electrostatics (\ref{1.22}).
This fact was verified on the 2D disk geometry, in the high-temperature
Debye-H\"uckel limit and at the exactly solvable coupling 
$\Gamma=\beta q^2=2$ of the OCP.
A progress towards the microscopic verification of formula (\ref{1.22})
was made in paper I.
There, the mapping of the 2D OCP, when $\Gamma$ is an even positive
integer, onto a discrete 1D anticommuting-field theory \cite{Samaj2} was
used for generating a sum rule for the charge structure function.
This sum rule comes from a specific unitary transformation of
anticommuting variables keeping a ``composite'' form of
the fermionic action.
For $\Lambda$ an elliptic domain, the sum rule confirms
microscopically the asymptotic formula (\ref{1.22}) and gives
a finite-size correction term to $\chi_{\Lambda}^i$ explicitly
in terms of boundary contributions.

The underlying sum rule derived for the 2D OCP seemed to be
closely related to the logarithmic nature of the 2D Coulomb potential.
We show in this paper that actually the sum rule is nothing but a 
direct consequence of a general principle: the total force 
acting on a system in thermal equilibrium is zero.
Using this principle, the sum rule is generalized to an arbitrary
$\nu$-dimensional Coulomb plasma.
As the result, for $\Lambda$ a $\nu$-dimensional ellipsoidal 
domain, the asymptotic formula (\ref{1.22}) for $\chi_{\Lambda}^i$ 
is reproduced and its leading finite-size correction
is obtained, in both canonical and grand canonical ensembles. 

The paper is organized as follows.
Section 2 is devoted to the derivation of the crucial sum rule
for an arbitrary $\nu$-dimensional Coulomb plasma.
Based on this sum rule, the splitting of the susceptibility
into its macroscopic part (\ref{1.22}) plus a corresponding
finite-size correction term is shown for $\nu$-dimensional
domains of ellipsoidal shape in Section 3.
Section 4 presents an analysis of the finite-size correction term, 
dependent on the particular ensemble.
The formalism is documented in Section 5 on the Debye-H\"uckel
limit.
The check on the exactly solvable 2D OCP at coupling $\Gamma=2$
has already been done in the previous paper I.
Concluding remarks are given in Section 6. 
 
\renewcommand{\theequation}{2.\arabic{equation}}
\setcounter{equation}{0}

\section{SUM RULES}
One of us has derived several sum rules for the 2D OCP
in paper I, using a mapping on a fermionic field theory. 
Actually, these sum rules are much more general.
In the present section, the generalization of some of these sum rules
is obtained by simple arguments about the balance of forces or torques.
   
Writing that the total force acting on the particles is zero, at
equilibrium, results into a sum rule relating their density 
$n({\bf r})$ and their charge density $\rho({\bf r})$:
\begin{equation} \label{2.1}
\rho_b \int_{\Lambda}{\rm d}^{\nu}r\, {\bf E}_b({\bf r}) \rho({\bf r})
- \frac{1}{\beta} \int_{\partial\Lambda}{\rm d}{\bf S}\, n({\bf r})=0
\end{equation}
where the first term in the l.h.s. is the force exerted by the 
background, and the second term is the force exerted by the walls.
${\rm d}{\bf S}\equiv {\rm d}S {\bf n}$ where ${\bf n}$ is the unit
vector normal to the surface element ${\rm d}S$ and directed
towards the exterior of $\Lambda$.
This is the generalization of Eqs. (56) of paper I. For simplicity, we
have assumed that the particle-wall interaction is a hard one, such that
the center of each particle feels a hard wall on $\partial\Lambda$. 

A similar sum rule is obtained by assuming that a particle of species
$\alpha_1$ is fixed at point ${\bf r}_1$, and writing that the total
force acting on the other particles vanishes. 
Now, the force that the fixed particle exerts on the other ones must 
also be included in the force balance, which reads
\begin{eqnarray}
\beta\rho_b\int_{\Lambda}{\rm d}^{\nu}r_2{\bf E}_b({\bf r}_2)
\sum_{\alpha_2} q_{\alpha_2} 
n_{\alpha_2\alpha_1}^{(2)}({\bf r}_2,{\bf r}_1)
-\int_{\partial\Lambda}{\rm d}{\bf S}_2 \sum_{\alpha_2}
n_{\alpha_2\alpha_1}^{(2)}({\bf r}_2,{\bf r}_1)& & \nonumber \\ 
+\beta\int_{\Lambda}{\rm d}^{\nu}r_2 \sum_{\alpha_2}
q_{\alpha_1}q_{\alpha_2}{\bf F}({\bf r}_2-{\bf r}_1)
n_{\alpha_2\alpha_1}^{(2)}({\bf r}_2,{\bf r}_1) = 0 & & \label{2.2}
\end{eqnarray}
where we have used that the density of particles of species $\alpha_2$
at ${\bf r}_2$, knowing that there is a particle of species $\alpha_1$
at ${\bf r}_1$, is 
$n_{\alpha_2\alpha_1}^{(2)}({\bf r}_2,{\bf r}_1)/
n_{\alpha_1}({\bf r}_1)$. 
If there are short-range interactions, they must be added to the 
definition (\ref{1.3}) of the Coulomb force $\bf F$.
Another form of Eq. (\ref{2.2}) can be obtained by using
the first BGY equation which can be written as 
\begin{eqnarray}
\nabla n_{\alpha_1}({\bf r}_1) & = & \beta\rho_b {\bf E}_b({\bf r}_1)
q_{\alpha_1}n_{\alpha_1}({\bf r}_1) \nonumber \\
& & + \beta\int_{\Lambda}{\rm d}^{\nu}r_2
\sum_{\alpha_2}q_{\alpha_1}q_{\alpha_2}{\bf F}({\bf r}_1-{\bf r}_2) 
n_{\alpha_2\alpha_1}^{(2)}({\bf r}_2,{\bf r}_1)  \label{2.3}
\end{eqnarray}
With regard to the equality ${\bf F}({\bf r}_1-{\bf r}_2)
= - {\bf F}({\bf r}_2-{\bf r}_1)$, using Eq. (\ref{2.3}) for the last
term in the l.h.s of Eq. (\ref{2.2}) gives
\begin{eqnarray}
\beta\rho_b\int_{\Lambda}{\rm d}^{\nu}r_2{\bf E}_b({\bf r}_2)
\left[ \sum_{\alpha_2} q_{\alpha_2}
n_{\alpha_2\alpha_1}^{(2)}({\bf r}_2,{\bf r}_1)
+q_{\alpha_1}n_{\alpha_1}({\bf r}_1)\delta({\bf r}_2-{\bf r}_1)\right]
& & \nonumber \\
-\int_{\partial\Lambda}{\rm d}{\bf S}_2 \sum_{\alpha_2}
n_{\alpha_2\alpha_1}^{(2)}({\bf r}_2,{\bf r}_1)
-{\bf \nabla}n_{\alpha_1}({\bf r}_1)=0 & & \label{2.4}
\end{eqnarray}
Finally, we multiply Eq. (\ref{2.4}) by $q_{\alpha_1}$ and sum on 
$\alpha_1$, we mutiply Eq. (\ref{2.1}) by $\beta\rho({\bf r}_1)$, and
we substract from each other the two resulting equations, with the
result
\begin{equation} \label{2.5}
\beta\rho_b\int_{\Lambda}{\rm d}^{\nu}r_2{\bf E}_b({\bf r}_2)
S({\bf r}_2,{\bf r}_1) = \nabla\rho({\bf r}_1) +
\int_{\partial\Lambda}{\rm d}{\bf S}_2\sum_{\alpha_1,\alpha_2}
q_{\alpha_1}U_{\alpha_2\alpha_1}({\bf r}_2,{\bf r}_1)
\end{equation}
This is the crucial sum rule which is the generalization of Eq. (60) 
of paper I.

Although we shall not need them in the following, let us mention that
another class of sum rules can be obtained from the balance of torques. 
For instance, in three dimensions, writing that the total
torque acting on the particles (due to both the background and the
walls) vanishes at equilibrium gives the sum rule
\begin{equation} \label{2.6}
\beta\rho_b\int_{\Lambda}{\rm d}^3r
[{\bf r}\times {\bf E}_b({\bf r})] \rho({\bf r})
+\int_{\partial\Lambda}[{\rm d}{\bf S}\times {\bf r}] n({\bf r}) = 0  
\end{equation}
This is the generalization of Eq. (41b) of paper I. 
If one particle is assumed to be fixed at some point, one obtains 
the torque analog of Eq. (\ref{2.5})
\begin{eqnarray}
\beta\rho_b\int_{\Lambda}{\rm d}^3r_2[{\bf r}_2\times 
{\bf E}_b({\bf r}_2)]S({\bf r}_2|{\bf r}_1)
& = & [{\bf r}_1\times \nabla] \rho({\bf r}_1) - \label{2.7} \\
& & \int_{\partial\Lambda}[{\rm d}{\bf S}_2\times {\bf r}_2]
\sum_{\alpha_1,\alpha_2} q_{\alpha_1}
U_{\alpha_2\alpha_1}({\bf r}_2,{\bf r}_1) \nonumber
\end{eqnarray}
This is the generalization of Eq. (45b) of paper I.

The sum rules (41a) and (45a) of paper I 
can also be generalized, following a method developed in 
refs. \cite{Choquard4} and \cite{Tellez}.
However, these generalizations will not be described here.

\renewcommand{\theequation}{3.\arabic{equation}}
\setcounter{equation}{0}

\section{DERIVATION OF THE SUSCEPTIBILITY}
Let $\Lambda$ be a $\nu$-dimensional ellipsoid in the reference
frame defined by the axes of the ellipsoid,
\begin{equation} \label{3.1}
\Lambda: \quad \quad
\sum_{i=1}^{\nu} \left( \frac{r^i}{R^i} \right)^2
\le 1
\end{equation}
In this reference frame both tensors $\chi_{\Lambda}$ and 
$T_{\Lambda}$ are diagonal.
For the domain shape under consideration, the depolarization
tensor $T_{\Lambda}$ is expressible as (\ref{1.23}) and 
independent of the point ${\bf r}\in \Lambda$, while
$\phi_b({\bf r})$ is invariant under the transformations
$r^i\to -r^i$.
This implies that 
\begin{equation} \label{3.2}
\phi_b({\bf r}) = {\rm const} - \frac{s_{\nu}}{2}
\sum_{i=1}^{\nu} T_{\Lambda}^i \left( r^i \right)^2
\end{equation}
The corresponding ${\bf E}_b({\bf r}) = - \nabla\phi({\bf r})$ reads
\begin{equation} \label{3.3}
{\bf E}_b({\bf r}) = s_{\nu} \sum_{i=1}^{\nu} T_{\Lambda}^i
r^i {\bf e}^i
\end{equation}
where ${\bf e}^i$ is the unit vector along the $i$th axis.
The components of $T_{\Lambda}$ for a 2D ellipse read
\begin{equation} \label{3.4}
T_{\Lambda}^1 = \frac{R^2}{R^1+R^2}, \quad \quad
T_{\Lambda}^2 = \frac{R^1}{R^1+R^2}
\end{equation}
The components of $T_{\Lambda}$ are more complicated functions
of $R^1, R^2, R^3$ for a 3D ellipsoid \cite{Landau}.
In the isotropic case $R^i=R$ of a $\nu$-dimensional sphere,
$T_{\Lambda}^i = 1/\nu$.

Inserting (\ref{3.3}) into the sum rule (\ref{2.5}), 
and defining the components
${\rm d}S_2^i = {\rm d}{\bf S}_2 \cdot {\bf e}^i$,
one gets for each component the equality
\begin{equation} \label{3.5}
\beta \rho_b s_{\nu} T_{\Lambda}^i \int_{\Lambda} 
{\rm d}^{\nu} r_2\, r_2^i S({\bf r}_2,{\bf r}_1) =
\frac{\partial}{\partial r_1^i} \rho({\bf r}_1)
+ \int_{\partial \Lambda} {\rm d}S_2^i \sum_{\alpha_1,\alpha_2}
q_{\alpha_1} U_{\alpha_2,\alpha_1}({\bf r}_2,{\bf r}_1)
\end{equation}
We multiply both sides of (\ref{3.5}) by $r_1^i$, then integrate
$\int_{\Lambda}{\rm d}^{\nu}r_1$ and use the definition (\ref{1.17})
of the dielectric susceptibility, for obtaining
\begin{equation} \label{3.6}
\rho_b s_{\nu} T_{\Lambda}^i \chi_{\Lambda}^i \vert \Lambda \vert
= \int_{\Lambda} {\rm d}^{\nu}r\, r^i \frac{\partial}{\partial r^i}
\rho({\bf r}) + \int_{\Lambda} {\rm d}^{\nu}r_1\, r_1^i
\int_{\partial \Lambda} {\rm d}S_2^i
\sum_{\alpha_1,\alpha_2} q_{\alpha_1} 
U_{\alpha_2\alpha_1}({\bf r}_2,{\bf r}_1)
\end{equation}
Simple algebra gives
\begin{eqnarray}
\int_{\Lambda} {\rm d}^{\nu}r\, 
r^i \frac{\partial}{\partial r^i} \rho({\bf r})
& = & \int_{\Lambda}{\rm d}^{\nu}r
\frac{\partial}{\partial r^i} \left[ r^i \rho({\bf r}) \right]
- \int_{\Lambda} {\rm d}^{\nu}r 
\left[ \rho({\bf r})+\rho_b-\rho_b \right] \nonumber \\
& = & \int_{\partial \Lambda} {\rm d}S^i\, r^i \rho({\bf r}) 
- \langle {\hat Q} \rangle
+ \rho_b \vert \Lambda \vert 
\label{3.7}
\end{eqnarray}
where 
\begin{equation} \label{3.8}
{\hat Q} = \int_{\Lambda} {\rm d}^{\nu}r 
\left[ {\hat\rho}({\bf r}) + \rho_b \right]
\end{equation}
is the microscopic total charge (including the fixed background charge) 
in the domain.
Provided that $\rho_b\ne 0$, Eq. (\ref{3.6}) can be thus rewritten
in the final form
\begin{equation} \label{3.9} 
\chi_{\Lambda}^i = \frac{1}{s_{\nu} T_{\Lambda}^i}
- \frac{1}{\rho_b s_{\nu} T_{\Lambda}^i} 
\left[ \frac{\langle {\hat Q}\rangle}{\vert \Lambda \vert}
- \frac{1}{\vert \Lambda\vert} \int_{\Lambda} {\rm d}^{\nu}r_1\,
r_1^i \int_{\partial\Lambda} {\rm d}S_2^i 
\langle {\hat\rho}({\bf r}_1){\hat n}({\bf r}_2) \rangle^{\rm T}
\right]
\end{equation}
This is the desired splitting of the susceptibility onto its
macroscopic part (\ref{1.22}) plus a finite-size correction term.

The formula (\ref{3.9}) can be further simplified in the isotropic
case of a $\nu$-dimensional spherical domain $\Lambda$ with
a radius $R^i=R$ and a volume 
$\vert\Lambda\vert = s_{\nu} R^{\nu}/\nu$.
Since now the components $\chi_{\Lambda}^i$ do not depend on $i$,
we can consider their common value
${\bar\chi}_{\Lambda} = \sum_{i=1}^{\nu} \chi_{\Lambda}^i/\nu$.
For the $\nu$-dimensional sphere it holds
\begin{equation} \label{3.10}
{\rm d}S_2^i = \frac{r_2^i}{R} {\rm d}S_2, \quad \quad
\sum_{i=1}^{\nu} r_1^i {\rm d}S_2^i = 
r_1 \cos\theta\, {\rm d}S_2 
\end{equation}
where $\theta$ is the angle between ${\bf r}_1$ and ${\bf r}_2$.
Since 
$\langle {\hat\rho}({\bf r}_1) {\hat n}({\bf r}_2) \rangle^{\rm T}$
depends on the orientations of ${\bf r}_1$ and ${\bf r}_2$ only
through their angle $\theta$, we can choose ${\bf r}_2$ along
the 1-axis and replace $\int_{\partial\Lambda}{\rm d}S_2$ 
by $s_{\nu}R^{\nu-1}$. 
Eq. (\ref{3.9}) takes the form
\begin{equation} \label{3.11} 
{\bar\chi}_{\Lambda}  =  \frac{\nu}{s_{\nu}}
- \frac{\nu}{\rho_b s_{\nu}} 
\left[ \frac{\langle {\hat Q}\rangle}{\vert \Lambda \vert} 
- \frac{1}{R} \int_{\Lambda} {\rm d}^{\nu}r\,
r^1 \langle {\hat\rho}({\bf r}){\hat n}({\bf R}) \rangle^{\rm T}
\right]
\end{equation}
where ${\bf R}=(R,0,\ldots,0)$.
It is sometimes convenient to express $r^1$ in the integral on
the r.h.s. of (\ref{3.11}) as $r^1 = R - (R-r^1)$ and in this way
to obtain an alternative ``boundary'' form of Eq. (\ref{3.11}),
\begin{equation} \label{3.12} 
{\bar\chi}_{\Lambda}  =  \frac{\nu}{s_{\nu}}
- \frac{\nu}{\rho_b s_{\nu}} \left[ 
\frac{\langle {\hat Q}\rangle}{\vert \Lambda \vert} 
- \langle {\hat Q} {\hat n}({\bf R}) \rangle^{\rm T}
+ \frac{1}{R} \int_{\Lambda} {\rm d}^{\nu}r\,
(R-r^1) \langle {\hat\rho}({\bf r}){\hat n}({\bf R}) \rangle^{\rm T}
\right] 
\end{equation}

The above formalism applies to the case $\rho_b\ne 0$, with no 
restriction on the use of canonical or grand-canonical ensembles.
When $\rho_b\to 0$, for the sake of simplicity we shall restrict
ourselves to the symmetric TCP in a $\nu$-dimensional
sphere and to only microscopic states such that the total charge 
of the system is equal to zero, ${\hat Q}=0$.
This is either the case of the canonical ensemble with imposed
charge neutrality, or the case of a restricted grand-canonical ensemble
when the fixed background of charge $-Nq$ is first neutralized
by $N$ opposite charges $+q$ and then $\pm q$ charges are added to 
the system in a variable number of neutral pairs \cite{Forrester}.
Under these conditions, relation (\ref{3.12}) reduces to
\begin{equation} \label{3.13}
{\bar\chi}_{\Lambda} = \frac{\nu}{s_{\nu}} -
\frac{\nu}{\rho_b s_{\nu}} \frac{1}{R} 
\int_{\Lambda} {\rm d}^{\nu}r \left( R-r^1 \right)
\langle {\hat\rho}({\bf r}) {\hat n}({\bf R}) 
\rangle^{\rm T}_{\rho_b}
\end{equation}
where the notation $\langle \cdots \rangle_{\rho_b}$ is used
to emphasize that the average is taken in presence of the background.
The background-charge density $\rho_b$ couples to particle coordinates
in the Boltzmann factor
$\exp[-\beta\rho_b \int_{\Lambda}{\rm d}^{\nu}r' \phi_b({\bf r}')
{\hat\rho}({\bf r}')]$, where $\phi_b$ is given in (\ref{3.2}).
In the limit $\rho_b\to 0$, the thermal average 
$\langle \cdots \rangle_{\rho_b}$ of a microscopic quantity can
be expanded around $\rho_b=0$, denoted simply as
$\langle \cdots \rangle$, using the linear response theory:
\begin{equation} \label{3.14}
\langle \cdots \rangle_{\rho_b} = \langle \cdots \rangle
- \beta \rho_b \int_{\Lambda} {\rm d}^{\nu}r' \phi_b({\bf r}')
\langle \cdots {\hat\rho}({\bf r}')\rangle^{\rm T}
+ O(\rho_b^2)
\end{equation}
Since, due to the $+\leftrightarrow -$ charge symmetry of the TCP,
$\langle {\hat\rho}({\bf r}) \rangle = 0$ and
$\langle {\hat\rho}({\bf r}) {\hat n}({\bf R}) \rangle = 0$
at any point ${\bf r}\in \Lambda$, relation (\ref{3.13}) can be 
rewritten in the $\rho_b\to 0$ limit as follows
\begin{eqnarray} 
{\bar\chi}_{\Lambda} & = & \frac{\nu}{s_{\nu}} - 
\frac{\beta}{2} \frac{1}{R}
\int_{\Lambda} {\rm d}^{\nu}r \left( R - r^1 \right)
\int_{\Lambda} {\rm d}^{\nu}r' \left( r' \right)^2
\nonumber \\ & & \times
\left[ \langle {\hat\rho}({\bf r}) {\hat\rho}({\bf r}')
{\hat n}({\bf R}) \rangle -
\langle {\hat\rho}({\bf r}) {\hat\rho}({\bf r}')\rangle
\langle {\hat n}({\bf R}) \rangle \right] \label{3.15}
\end{eqnarray}
We see that for the TCP with no background, three-body densities
enter the finite-size contribution. 

\renewcommand{\theequation}{4.\arabic{equation}}
\setcounter{equation}{0}

\section{NON-EQUIVALENCE OF ENSEMBLES}
Although the macroscopic result for the $\nu$-dimensional sphere
${\bar\chi}_{\Lambda} \sim \nu/s_{\nu}$ is the same in both the
canonical and grand-canonical ensembles, the finite-size correction 
term in (\ref{3.12}) is ensemble-dependent.

\subsection{Canonical Ensemble}
In the canonical ensemble, the microscopic total charge is fixed, 
${\hat Q}=Q$.
Let us analyze term by term the finite-size corrections
appearing on the r.h.s. of Eq. (\ref{3.12}).

If there is some excess charge in the domain $\Lambda$, due to
the electrostatic repulsion it has tendency to move to the
domain boundary $\partial \Lambda$ and to create there a macroscopic
surface charge density $\sigma = Q/\vert \partial\Lambda\vert$.
We note that, as a consequence, 
$\langle {\hat Q}\rangle/\vert\Lambda\vert = \nu\sigma/R$,
and it is reasonable to assume that $\sigma$ is finite. 
The other thermal averages in (\ref{3.12}) are assumed to be
taken for a fixed $\sigma$.

Since the microscopic total charge does not fluctuate,
$\langle {\hat Q}{\hat n}({\bf R}) \rangle^{\rm T} = 0$.

One has to be cautious when identifying the $R\to\infty$ limit
of the dipole moment in the last term with its flat hard-wall
counterpart:
owing to a slow power-law decay of the correlations along a plain 
hard wall \cite{Jancovici1}-\cite{Jancovici4}, the limit cannot 
be freely interchanged with the integration.
In particular, let us consider in $\nu$-dimensions a semi-infinite 
Coulomb plasma which occupies the half-space $x>0$; 
we denote by ${\bf y}$ the set of remaining $(\nu-1)$ coordinates 
normal to $x$.
The plane at $x=0$ is charged with the uniform surface
charge density $\sigma$.
It is shown in Appendix that in dimensions $\nu=2,3$ 
the $R\to\infty$ limit of the considered sphere dipole moment 
is related to the corresponding flat dipole moment as follows:
\begin{equation} \label{4.1}
\lim_{R\to\infty} \int_{\Lambda} {\rm d}^{\nu}r\,
(R-r^1) \langle {\hat\rho}({\bf r}){\hat n}({\bf R}) \rangle^{\rm T}
= 2 \int_0^{\infty} {\rm d}x\, x \int {\rm d}{\bf y}
\langle {\hat\rho}(x,{\bf y}) {\hat n}(0,{\bf 0}) \rangle^{\rm T}
\end{equation}
The factor 2 in this equation was first observed in paper I
for the case of the 2D OCP at coupling $\Gamma=2$.
Its temperature-independence is also checked in the Debye-H\"uckel 
limit (see the next section).

We conclude that in $\nu=2,3$ dimensions the formula for the 
dielectric susceptibility tensor of the Coulomb conductor, 
evaluated in the canonical ensemble up to the leading $1/R$ 
finite-size correction term, reads
\begin{equation} \label{4.2}
{\bar\chi}_{\Lambda} \sim \frac{\nu}{s_{\nu}} -
\frac{\nu}{\rho_b s_{\nu}} \frac{1}{R} \left[
\nu \sigma + 2 \int_0^{\infty} {\rm d}x\, x \int {\rm d}{\bf y}
\langle {\hat\rho}(x,{\bf y}) {\hat n}(0,{\bf 0}) \rangle^{\rm T}
\right]
\end{equation}
This result can be readily extended to the $\rho_b\to 0$
limit of the symmetric TCP with ${\hat Q}=0$ (and, consequently,
$\sigma=0$), discussed at the end of the previous section.
Using for the truncated correlation in (\ref{4.2}) the linear response 
(\ref{3.14}), now in the half-space geometry with 
$\phi_b({\bf r}') = - s_{\nu}(x')^2/2$, one arrives at
\begin{eqnarray} 
{\bar\chi}_{\Lambda} & \sim & \frac{\nu}{s_{\nu}}
- \nu \beta \frac{1}{R} \int_0^{\infty} {\rm d}x\, x \int {\rm d}{\bf y}
\int_0^{\infty} {\rm d}x'\, (x')^2 \int {\rm d}{\bf y}'
\nonumber \\ & &
\left[ \langle {\hat\rho}(x,{\bf y}) {\hat\rho}(x',{\bf y}')
{\hat n}(0,{\bf 0}) \rangle - 
\langle {\hat\rho}(x,{\bf y}) {\hat\rho}(x',{\bf y}') \rangle
\langle {\hat n}(0,{\bf 0}) \rangle \right] \label{4.3}
\end{eqnarray}
Although the finite-size analysis was made for the $\nu=2,3$
spherical geometries, it can be simply generalized via 
Eq. (\ref{3.9}) to an arbitrary $\nu$-dimensional ellipsoid:
the leading correction term is still of the order of 1 over
the characteristic length of the domain. 

\subsection{Grand Canonical Ensemble}
The grand-canonical analysis of the finite-size corrections
in (\ref{3.12}) fundamentally depends on the dimension.

\noindent {\it Two Dimensions}. In the grand canonical ensemble,
necessarily the total charge $\hat{Q}$ vanishes and does not fluctuate 
\cite{Jancovici6} (except in a very special case not discussed here). 
This is because bringing a charged particle into the
system from a reservoir at infinity, with a hole left in the reservoir,
would cost an infinite energy, and this cannot be achieved with finite
fugacities. Thus, the terms $\langle\hat{Q}\rangle$ and 
$\langle {\hat Q}{\hat n}({\bf R})\rangle^{\rm T}$ vanish in
(\ref{3.11}) and (\ref{3.12}). Furthermore, (\ref{4.1}) and
(\ref{4.3}) are still valid.

\noindent {\it Three Dimensions}. In the grand canonical ensemble, for a
finite system, $\langle\hat{Q}\rangle$ is determined by the fugacities
and does not vanish, except for special adjustments of these
fugacities. However, $\langle\hat{Q}\rangle$ is at most of order
$R$. Indeed, when the sphere already carries a charge $Q$, the work
required from bringing one more particle of charge $q$ into the system
from the reservoir has an electric part $qQ/R$. Therefore, with finite
chemical potentials, $qQ/R$ has to be finite.
 
The total charge does fluctuate, with a variance such that
$\beta\langle\hat{Q}^2\rangle^{\rm T}=R$ in the large-$R$ limit, and 
the term $\langle {\hat Q}{\hat n}({\bf R}) \rangle^{\rm T}$ in  
(\ref{3.12}) does not vanish, and is of order $1/R$ as shown below.

Indeed, considering for simplicity the case of the OCP in a 3D sphere of
radius R, $\langle {\hat Q}{\hat n}({\bf R})\rangle^{\rm T}$ is
proportional to the total charge on the sphere when one of the particles
of charge $q$ is fixed on the surface. Macroscopic electrostatics says
that, when a point charge $q$ is at distance $r\geq R$ from the center
of a grounded sphere, it induces on it a surface charge $q'=-(R/r)q$.
Thus, the total charge $q+q'$ vanishes if $r=R$. However, actually, the
``surface'' charge has some microscopic thickness $\lambda$ of the
order of the charge correlation length, and it is
better to describe approximately the configuration of a particle fixed
on the surface as a particle at distance $R$ from the center of a sphere
of radius $R-\lambda$. 
Thus $q'=-[(R-\lambda)/R]q$, the total charge $q+q'$ is of order 
$q\lambda/R$, and $\langle {\hat Q} {\hat n}(R) \rangle^{\rm T}$
is expected to be of order $\rho \lambda/R$. 

Finally, in (\ref{3.11}) and (\ref{3.12}), in the large-$R$ limit, the
term $\langle\hat{Q}\rangle/|\Lambda|$ is at most of order $1/R^2$ an
can be discarded. But the term 
$\langle {\hat Q}{\hat n}({\bf R})\rangle^{\rm T}$ gives to
(\ref{3.12}) a contribution of order $1/R$, like the dipole integral,
and both should be kept in the leading finite-size
correction. 
As to (\ref{4.1}) and (\ref{4.3}), they are still valid.
 
\renewcommand{\theequation}{5.\arabic{equation}}
\setcounter{equation}{0}

\section{DEBYE-H\"{U}CKEL THEORY}
The formulas (\ref{3.11}) and (\ref{4.1}) will now be tested in the
weak-coupling limit, which is described by the Debye-H\"{u}ckel theory,
for the general system of $M$ species of point particles plus a
background, in two or three dimensions.

\subsection{General Formalism} 
A consistent way of deriving the Debye-H\"{u}ckel theory for a finite
system is to start with the renormalized Mayer diagrammatic expansion 
(which is reviewed, for instance, in refs. \cite{Kalinay} and 
\cite{Jancovici5}), in the grand canonical ensemble, and to make 
a topological reduction, replacing the fugacities by the densities. 
The weak-coupling limit for the correlation functions is obtained 
by resumming the chain diagrams with the densities taken as constants 
$n_{\alpha}$ (taking into account their position-dependence near 
the boundary $\partial\Lambda$ would give corrections of higher order). 
This is equivalent to writing the Ornstein-Zernicke equations with 
the direct correlation functions replaced by $-\beta$ times 
the corresponding interaction potential:
\begin{eqnarray}
h_{\alpha_1\alpha_2}({\bf r}_1,{\bf r}_2) & = &
-\beta q_{\alpha_1}q_{\alpha_2}v(|{\bf r}_1-{\bf r}_2|) \label{5.1} \\
& & + \sum_{\alpha_3}\int_{\Lambda}{\rm d}^{\nu}r_3
[-\beta q_{\alpha_1}q_{\alpha_3}v(|{\bf r}_1-{\bf r}_3|)]n_{\alpha_3}
h_{\alpha_3\alpha_2}({\bf r}_3,{\bf r}_2) \nonumber
\end{eqnarray}
where the correlation functions $h$ are related to the Ursell functions
by $U_{\alpha_1\alpha_2}({\bf r}_1,{\bf r}_2)=n_{\alpha_1}n_{\alpha_2}
h_{\alpha_1\alpha_2}({\bf r}_1,{\bf r}_2)$. 
The set (\ref{5.1}) of $M^2$ coupled equations can be transformed 
into one equation. 
Indeed, let us make the ansatz that the solution is of the form
\begin{equation} \label{5.2}
h_{\alpha_1\alpha_2}({\bf r}_1,{\bf r}_2)=
-\beta q_{\alpha_1}q_{\alpha_2}G({\bf r}_1,{\bf r}_2) 
\end{equation}
Using (\ref{5.2}) in (\ref{5.1}) one does check that these
Ornstein-Zernicke equations are satisfied provided that $G$ obeys the
integral equation 
\begin{equation} \label{5.3}
G({\bf r}_1,{\bf r}_2)=v(|{\bf r}_1-{\bf r}_2|)
-\frac{\kappa^2}{s_{\nu}}\int_{\Lambda}{\rm d}^{\nu}r_3
v(|{\bf r}_1-{\bf r}_3|)G({\bf r}_3,{\bf r}_2) 
\end{equation}
where $\kappa^2=s_{\nu}\beta\sum_{\alpha}n_{\alpha}q_{\alpha}^2$; the
Debye length is $1/\kappa$. 
Using (\ref{5.2}) one finds
\begin{equation} \label{5.4}
\langle\hat{\rho}({\bf r}_1)\hat{n}({\bf r}_2)\rangle^{\rm T}=
-\frac{\rho\kappa^2}{s_{\nu}}G({\bf r}_1,{\bf r}_2) 
+\rho\delta ({\bf r}_1-{\bf r}_2)     
\end{equation}
and
\begin{equation} \label{5.5}
S({\bf r}_1,{\bf r}_2)\equiv
\langle\hat{\rho}({\bf r}_1)\hat{\rho}({\bf r}_2)\rangle^{\rm T}=
-\frac{1}{\beta}\left(\frac{\kappa^2}{s_{\nu}}\right)^2
G({\bf r}_1,{\bf r}_2)+\frac{\kappa^2}{\beta s_{\nu}}
\delta ({\bf r}_1-{\bf r}_2)   
\end{equation}

The integral equation (\ref{5.3}) for $G$ can be transformed into a 
differential equation by taking the Laplacian with respect to 
${\bf r}_1$. One obtains the usual Debye-H\"{u}ckel equation for the
screened Coulomb potential $G$
\begin{equation} \label{5.6}
[\Delta_1-\kappa^2] G({\bf r}_1,{\bf r}_2)=
-s_{\nu}\delta({\bf r}_1-{\bf r}_2)  
\end{equation}
However, in a finite system, the differential equation (\ref{5.6}) must
be supplemented by boundary conditions. 
In the present approach, these boundary conditions are provided 
by the integral equation (\ref{5.3}).

It has been seen in Section 3 that, in general, when there is a
background, the finite-size correction to the susceptibility 
can be expressed in terms of the two-body correlation appearing 
in (\ref{3.11}), while, in the limit of no background, one obtains the
more complicated expression (\ref{3.15}) in terms of a three-body
correlation. The Debye-H\"{u}ckel theory has the very special feature
that this complication does not arise. Indeed, since $\rho=-\rho_b$,
one sees in (\ref{5.4}) that 
$\langle\hat{\rho}({\bf r}_1)\hat{n}({\bf r}_2)\rangle^{\rm T}/\rho_b$
is expressed in terms of the two-body function $G$ even in the limit
$\rho_b\to 0$. Furthermore $\langle\hat{Q}\rangle =0$. Therefore
(\ref{3.11}) still involves only a two-body correlation in this limit
$\rho_b\to 0$.
  
\subsection{2D Disk}
In an infinite plane, (\ref{5.6}) gives  
$G({\bf r}_1,{\bf r}_2) = K_0(\kappa|{\bf r}_1-{\bf r}_2|)$, 
where $K_0$ is a modified Bessel function. 
In a finite disk of radius $R$, the solution is of the 
form \cite{Choquard3}
\begin{equation} \label{5.7}
G({\bf r}_1,{\bf r}_2)=\sum_{\ell=0}^{\infty}[I_{\ell}(s_<)K_{\ell}(s_>)
+a_{\ell}I_{\ell}(s_1)I_{\ell}(s_2)]\mu_{\ell}\cos\ell\theta
\end{equation}
where $s_{1,2}=\kappa r_{1,2}$, $s_<$ and $s_>$ are the smallest and the
largest, respectively, of $s_1$ and $s_2$, $I_{\ell}$ and $K_{\ell}$ are
modified Bessel functions, and $a_{\ell}$ a coefficient to be
determined; $\mu_{\ell}$ is the Neumann factor $\mu_0=1$, $\mu_{\ell}=2$
for $\ell\geq 1$. 
In the square bracket of (\ref{5.7}) the first term
corresponds to an expansion of $K_0(\kappa|{\bf r}_1-{\bf r}_2|)$, while
the second term corresponds to the general symmetric solution of
(\ref{5.6}) without the r.h.s. $\delta$ term.
 
The determination of $a_0$ from the integral equation (\ref{5.3}) 
has been discussed in ref. \cite{Jancovici6}, where it has been 
argued that the length scale $r_0$ in the 2D Coulomb potential $v$ 
must be made infinite at the end of the calculation. 
The result is $a_0=K_1(Z)/I_1(Z)$, where $Z=\kappa R$.
  
For determining $a_{\ell}$ when $\ell\geq 1$, we consider the integral
equation (\ref{5.3}), and use for $G$ the expansion (\ref{5.7}) and
for $v$ the expansion
\begin{equation} \label{5.8}
v(|{\bf r}_1-{\bf r}_2|)=-\ln\frac{|{\bf r}_1-{\bf r}_2|}{r_0}=
-\ln\frac{r_>}{r_0}+\sum_{\ell =1}^{\infty}\frac{1}{\ell}
\left(\frac{r_<}{r_>}\right)^{\ell}\cos\ell(\theta_2-\theta_1)
\end{equation}
In the angular integral on $\theta_3$, only the terms involving the same
$\ell$ in the two expansions (\ref{5.7}) and (\ref{5.8}) survive. 
In terms of the square bracket in (\ref{5.7}), i.e.
\begin{equation} \label{5.9}
G_{\ell} (r_1,r_2)\equiv 
I_{\ell}(s_<)K_{\ell}(s_>)+a_{\ell}I_{\ell}(s_1)I_{\ell}(s_2) 
\end{equation}
one obtains, when $r_1>r_2$,
\begin{eqnarray} 
2G_\ell (r_1,r_2)&=&\frac{1}{\ell}\left(\frac{r_2}{r_1}\right)^{\ell}
-\kappa^2\int_0^{r_1}{\rm d}r_3r_3
\frac{1}{\ell}\left(\frac{r_3}{r_1}\right)^{\ell}G_\ell (r_3,r_2)
\nonumber \\
&-&\kappa^2\int_{r_1}^R{\rm d}r_3r_3\frac{1}{\ell}
\left(\frac{r_1}{r_3}\right)^{\ell}G_\ell (r_3,r_2),\;\;\;\ell\geq 1
\label{5.10}  
\end{eqnarray}
One could solve (\ref{5.10}). 
However, it is simpler to remark that it implies
$\partial G_{\ell}(r_1,r_2)/\partial r_1|_{r_1=R} 
= -(\ell /R)G_{\ell}(R,r_2)$.
Therefore, using the definition (\ref{5.9}) for $G_{\ell}(R,r_2)$ gives
\begin{equation} \label{5.11}
K'_{\ell}(Z)+a_{\ell}I'_{\ell}(Z)=-\frac{\ell}{Z}
[K_{\ell}(Z)+a_{\ell}I_{\ell}(Z)],\;\;\;\ell\geq 1 
\end{equation}
a relation that Choquard et al.\cite{Choquard3} have obtained by
another method, involving a continuation of (\ref{5.6}) outside the
disk; that method led to some ambiguity for determining $a_0$. 
From (\ref{5.11}), using simple relations obeyed by the Bessel
functions, one obtains 
\begin{equation} \label{5.12}
a_{\ell}=\frac{K_{\ell -1}(Z)}{I_{\ell -1}(Z)}  
\end{equation} 
This final equation turns out to be valid for all $\ell$, including
$\ell =0$.

Using (\ref{5.7}) and (\ref{5.12}) in (\ref{5.2}), one can easily
check the perfect screening expected in two dimensions, even in the 
grand canonical ensemble: the charge in the cloud around a particle of
charge $q_{\alpha_2}$ is $-q_{\alpha_2}$,
\begin{equation} \label{5.13}
\int_{\Lambda}{\rm d}^2r_1\sum_{\alpha_1}q_{\alpha_1}n_{\alpha_1}
h_{\alpha_1\alpha_2}({\bf r}_1,{\bf r}_2)=-q_{\alpha_2}   
\end{equation}

We now turn to the dielectric susceptibility. From its definition
(\ref{1.17}), using the present $S({\bf r}_1,{\bf r}_2)$,
one obtains \cite{Choquard3}
\begin{equation} \label{5.14}
\bar{\chi}_{\Lambda}=\frac{1}{\pi}\left[1-\frac{2I_1(\kappa R)}
{\kappa RI_0(\kappa R)}\right]  
\end{equation}
Alternatively, one can use the general method of the present paper and
check the expression (\ref{3.11}). 
Here $\langle \hat{Q}\rangle =0$, and
$\rho_b=-\rho$. Let $-D_{\rm disk}$ be the dipole moment defined as the 
integral in (\ref{3.11}) (we call this dipole moment $-D$ rather than
$D$ for using the same notation as in the Appendix). 
Only the part $\ell =1$ of $G$ contributes to this integral. 
Using $\langle\hat{\rho}({\bf r})\hat{n}({\bf R})\rangle^{\rm T}$ from
(\ref{5.4}) and $G_1$ from (\ref{5.9}) with $a_1$ from (\ref{5.11})
gives, after simple manipulations on the Bessel functions,
\begin{equation} \label{5.15}
D_{\rm disk}=-\frac{2\rho I_1(\kappa R)}{I_0(\kappa R)} 
\end{equation}
It should be remarked that, since in 2D 
$\langle \hat{Q}\hat{n}({\bf R)}\rangle^{\rm T} = 0$, 
$D_{\rm disk}$ is also the integral in (\ref{3.12}). 
Using (\ref{5.15}) in (\ref{3.11}) or (\ref{3.12}), 
one retrieves the same $\bar{\chi}_{\Lambda}$ as in
(\ref{5.14}). 
In the large-$R$ limit, in (\ref{5.14}) 
$I_1(\kappa R)/I_0(\kappa R)\rightarrow 1$ and one sees
that the correction term is indeed of order $1/R$:  
\begin{equation} \label{5.16}
\bar{\chi}_{\Lambda}\sim\frac{1}{\pi}-\frac{2}{\pi\kappa R} 
\end{equation}

The dipole moment $D_{\rm flat}$ for a flat wall, in the 2D 
Debye-H\"{u}ckel theory, has been computed in \cite{Jancovici3}. 
It can be checked that, in the limit $R\rightarrow\infty$, 
$D_{\rm disk}$ does have twice the value found for $D_{\rm flat}$.  

\subsection{3D Sphere}
In infinite space, (\ref{5.6}) gives $G({\bf r}_1,{\bf r}_2)=
\exp(-\kappa |{\bf r}_1-{\bf r}_2|)/|{\bf r}_1-{\bf r}_2|$. 
In a finite sphere of radius $R$, the same considerations as in 2D now
give \cite{Choquard3}
\begin{equation} \label{5.17}
G({\bf r}_1,{\bf r}_2)=\sum_{\ell=0}^{\infty}\frac{2\ell +1}
{\sqrt{r_1r_2}}\left[I_{\ell +\frac{1}{2}}(s_<)
K_{\ell +\frac{1}{2}}(s_>) + 
b_{\ell}I_{\ell +\frac{1}{2}}(s_1)I_{\ell +\frac{1}{2}}(s_2)\right]
P_{\ell}(\cos\theta)   
\end{equation}
where $P_{\ell}$ is a Legendre polynomial. 
As in 2D, the coefficients $b_{\ell}$ can be determined by using 
the integral equation (\ref{5.3}), with the same result as 
in ref. \cite{Choquard3}:
\begin{equation} \label{5.18}
b_{\ell}=\frac{K_{\ell -\frac{1}{2}}(Z)}{I_{\ell -\frac{1}{2}}(Z)}
\end{equation}
With our method, there is no special problem or ambiguity with the case 
$\ell=0$.

As expected, there is no perfect screening, since the starting point was
the grand canonical ensemble. 
Using (\ref{5.17}) and (\ref{5.18}) in (\ref{5.2}) gives
\begin{equation} \label{5.19}
\int_{\Lambda}{\rm d}^3r_1\sum_{\alpha_1}q_{\alpha_1}n_{\alpha_1}
h_{\alpha_1\alpha_2}({\bf r}_1,{\bf r}_2)=-q_{\alpha_2}
\left[1-\frac{\sinh\kappa r_2}{\kappa r_2\cosh\kappa R}\right]
\end{equation}  
rather than $-q_{\alpha_2}$.

The dielectric susceptibility, computed from its definition (\ref{1.17})
is found to be \cite{Choquard3}
\begin{equation} \label{5.20}
\bar{\chi}_{\Lambda}=\frac{3}{4\pi}\left[1-\frac{3I_{\frac{3}{2}}
(\kappa R)}{\kappa RI_{\frac{1}{2}}(\kappa R)}\right]
{ \atop \stackrel{\sim}{\scriptstyle{R\to\infty}} }
\frac{3}{4\pi}\left[1-\frac{3}{\kappa R}\right]
\end{equation}
Alternatively, one can use the general method of the present paper. 
Again $\langle \hat{Q}\rangle =0$, $\rho_b=-\rho$, and  
only the part $\ell=1$ of $G$ contributes to the integral in
(\ref{3.11}). 
One retrieves for the susceptibility the result (\ref{5.20}). 

It should be noted that, in 3D, $\hat{Q}$ fluctuates and  
$\langle \hat{Q}\hat{n}({\bf R)}\rangle^{\rm T}\neq 0$. One finds
\begin{equation} \label{5.21}
\langle \hat{Q}\hat{n}({\bf R)}\rangle^{\rm T}
=\rho\frac{I_{\frac{1}{2}}(\kappa R)}
{\kappa R I_{-\frac{1}{2}}(\kappa R)}
{ \atop \stackrel{\sim}{\scriptstyle{R\to\infty}} }
\frac{\rho}{\kappa R}   
\end{equation}
in agreement with the qualitative estimate of Section 4.2. 
Therefore, with $D_{\rm sph}$ defined as the integral in (\ref{3.12}),
the equivalence of (\ref{3.11}) and (\ref{3.12}) gives
\begin{equation} \label{5.22}
D_{\rm sph}
{ \atop \stackrel{\sim}{\scriptstyle{R\to\infty}} }
-\frac{2\rho}{\kappa} 
\end{equation}
and the finite-size correction to $\bar{\chi}_{\Lambda}$ can be
decomposed as
\begin{equation} \label{5.23}
\bar{\chi}_{\Lambda}
{ \atop \stackrel{\sim}{\scriptstyle{R\to\infty}} }
\frac{3}{4\pi}\left[1-\frac{1}{\kappa R}-\frac{2}{\kappa R}\right]
\end{equation}
where the term $1/\kappa R$ is the contribution from 
$\langle \hat{Q}\hat{n}({\bf R)}\rangle^{\rm T}$ and the term
$2/\kappa R$ is the contribution from the dipole moment $D_{\rm sph}$
seen fom the boundary. Again, in the limit $R\rightarrow\infty$, 
$D_{\rm sph}$ does have twice the value found for a flat
wall \cite{Jancovici3} in the Debye-H\"{u}ckel theory.

\renewcommand{\theequation}{6.\arabic{equation}}
\setcounter{equation}{0}

\section{CONCLUSION}
Macroscopic electrostatics predicts a shape-dependent value for the
dielectric susceptibility of a conductor (the response, sometimes called
polarizability, to a uniform applied electric field). In the present
paper, it has been shown that classical (i.e. non-quantum) equilibrium 
statistical mechanics of a large class of microscopic models results
into a dielectric susceptibility which is the sum of the macroscopic
value plus an explicit finite-size correction. Thus, the limits of
validity of macroscopic electrostatics are clearly exhibited. 

The basis for the microscopic derivation only is that the total force
acting on a system vanishes at equilibrium. It is quite surprising
that such a simple statement is sufficient, and the reason for that
still is an open problem.

Our approach deals with models of Coulomb systems made of charged
particles embedded in a uniformly charged background. The case
of no background is dealt with as a limiting case. It seems that our
method cannot be used for directly starting with a system without a
background.

Classical statistical mechanics has been used. It gives an acceptable
phenomenological description of some systems such as electrolytes or
molten salts. We have not attempted to deal with a more fundamental 
description of real matter based on quantum statistical mechanics of
point charges.

\renewcommand{\theequation}{A.\arabic{equation}}
\setcounter{equation}{0}

\section*{APPENDIX: DIPOLE MOMENTS}
We briefly summarize known facts about the large-distance behavior
of particle correlations along a plain, possibly homogeneously
charged, hard-wall in $\nu=2,3$ dimensions.
Let us first review the case of a semi-infinite Coulomb plasma
which occupies the half-space $x>0$; ${\bf y}$ denotes the set
of $(\nu-1)$ coordinates normal to the $x$-axis.
According to ref. \cite{Jancovici3}, for the charge-density correlator 
one expects an asymptotic power-law behavior along the boundary of type
\begin{equation} \label{A.1}
\langle {\hat\rho}(x,{\bf y}){\hat n}(x',{\bf 0}) \rangle^{\rm T}
\sim \frac{g_{\nu}(x,x')}{\vert {\bf y}\vert^{\nu}}, \quad \quad
\vert {\bf y} \vert \to \infty
\end{equation}
where $g_{\nu}(x,x')$, which as a function of $x$ and $x'$ has a fast 
decay away from the wall, obeys the relation
\begin{equation} \label{A.2}
\frac{s_{\nu}}{2} \int_0^{\infty} {\rm d}x\, g_{\nu}(x,x')
= \int_0^{\infty} {\rm d}x\, x \int {\rm d}{\bf y} 
\langle {\hat\rho}(x,{\bf y}) {\hat n}(x',{\bf 0}) \rangle^{\rm T},
\quad \quad \nu=2,3
\end{equation}
valid for any $x'\ge 0$.
A behavior of type (\ref{A.1}) at large distances was observed 
also in the large-$R$ limit of the 
$\nu$-dimensional sphere \cite{Choquard2,Jancovici4}.
For two points ${\bf r}$ and ${\bf r}'$ inside the sphere, 
it is only necessary to identify $x$ and $x'$ with 
the corresponding point distances from the sphere surface 
and $\vert{\bf y}\vert$ with the Euclidean distance (chord) of 
the point projections onto the sphere surface:
\begin{equation} \label{A.3}
x=R-r, x'=R-r';\quad \quad
\vert {\bf y} \vert = 2 R \sin(\theta/2)
\end{equation}
where $\theta$ is the angle between ${\bf r}$ and ${\bf r}'$.
At small distances, an infinitesimal deformation of a flat 
boundary towards the sphere has a negligible effect 
on the correlations.
We can therefore write, on both microscopic and macroscopic scales,
that, as the radius of the sphere $R\to\infty$,
\begin{equation} \label{A.4}
\langle {\hat\rho}({\bf r}){\hat n}({\bf r}') \rangle^{\rm T}
\big\vert_{\rm sphere} \sim
\langle {\hat\rho}(x,{\bf y}){\hat n}(x',{\bf 0}) \rangle^{\rm T}
\big\vert_{\rm flat} 
\end{equation} 

In the dipole integral on the r.h.s. of (\ref{3.12}), the correlator 
of interest is taken at the point ${\bf r}'={\bf R}$ fixed at
the boundary, which corresponds to $x'=0$ in (\ref{A.3}).
To simplify the notation, we define
\begin{equation} \label{A.5}
\psi_{\rm sph}(x,\theta) =
\langle {\hat\rho}({\bf r}){\hat n}({\bf R}) \rangle^{\rm T}
\big\vert_{\rm sphere}, \quad \quad
\psi_{\rm flat}(x,\vert {\bf y} \vert) =
\langle {\hat\rho}(x,{\bf y}){\hat n}(0,{\bf 0}) \rangle^{\rm T}
\big\vert_{\rm flat}
\end{equation}
Within the identification (\ref{A.3}) with $x'=0$, the asymptotic 
$R\to \infty$ equivalence (\ref{A.4}) now takes the form
\begin{equation} \label{A.6}
\psi_{\rm sph}(x,\theta) \sim \psi_{\rm flat}(x,\vert {\bf y} \vert);
\quad \quad \vert {\bf y} \vert = 2 R \sin(\theta/2)
\end{equation}
Our task is to relate the $R\to\infty$ limit of the sphere
dipole moment $D_{\rm sph}$ seen from the boundary and the flat dipole
moment $D_{\rm flat}$, defined as follows
\begin{eqnarray}
D_{\rm sph} & = &
\int_{\Lambda} {\rm d}^{\nu}r (R-r^1) 
\psi_{\rm sph}(R-r,\theta) \label{A.7} \\
D_{\rm flat} & = &
\int_0^{\infty} {\rm d}x\, x \int {\rm d}{\bf y}\,
\psi_{\rm flat}(x,\vert {\bf y} \vert) \label{A.8} 
\end{eqnarray}
Because of slight differences, the derivations of the relation are made 
separately for 2D (with notation ``disk'' instead of ``sph'') and 3D.

\subsection*{2D Disk}
Using the substitution $x=R-r$ and writing
$r^1 = (R-x)[1-2\sin^2(\theta/2)]$, 
the disk dipole moment (\ref{A.7}) is expressible as
\begin{eqnarray}
D_{\rm disk} & = & \int_0^R {\rm d}x\, x (R-x) 
\int_{-\pi}^{\pi} {\rm d}\theta\, \psi_{\rm disk}(x,\theta) 
\nonumber \\
& & + 2 \int_0^R {\rm d}x\, (R-x)^2 \int_{-\pi}^{\pi} {\rm d}\theta\, 
\sin^2(\theta/2) \psi_{\rm disk}(x,\theta) 
\label{A.9}
\end{eqnarray}
In the large-$R$ limit, we make use of the transformation (\ref{A.6})
to get $D_{\rm disk} = I_1 + I_2$, where
\begin{eqnarray}
I_1 & = & \int_0^R {\rm d}x\, x \left( 1 - \frac{x}{R} \right)
\int_{-2R}^{2R} \frac{{\rm d}y}{\sqrt{1 - \frac{y^2}{4 R^2}}}
\psi_{\rm flat}(x,y) \label{A.10} \\
I_2 & = & 2 \int_0^R {\rm d}x\, R \left( 1- \frac{x}{R} \right)^2
\int_{-2R}^{2R} \frac{{\rm d}y}{\sqrt{1 - \frac{y^2}{4 R^2}}}
\frac{y^2}{4 R^2} \psi_{\rm flat}(x,y) \label{A.11}
\end{eqnarray}
Since $\psi_{\rm flat}(x,y)$ as a function of $x$ has a fast decay
away from the wall, the $x/R$ terms in $I_1$ and $I_2$ can be neglected 
in comparison with the unity when $R\to\infty$.
After simple algebra, one finds for $I_1$
\begin{equation} \label{A.12}
\lim_{R\to\infty} I_1 = D_{\rm flat} + \int_{0}^{\infty} {\rm d}x\, x
\lim_{R\to\infty} 2 R \int_{-1}^1 {\rm d}t 
\left( \frac{1}{\sqrt{1-t^2}}-1 \right) \psi_{\rm flat}(x,2tR)
\end{equation} 
Considering $\psi_{\rm flat}(x,2tR) \sim g_2(x,0)/(2tR)^2$ 
implies a converging integral over $t$,
so that $\lim_{R\to\infty} I_1 = D_{\rm flat}$.
As concerns the second integral $I_2$, it can be analogously
written as
\begin{equation} \label{A.13}
\lim_{R\to\infty} I_2 = \int_0^{\infty} {\rm d}x
\lim_{R\to\infty} (2R)^2 \int_{-1}^1 \frac{{\rm d}t}{\sqrt{1-t^2}}
t^2 \psi_{\rm flat}(x,2tR)
\end{equation}
As above, we consider the leading asymptotic behavior
of $\psi_{\rm flat}(x,2tR)$, with the result
\begin{equation} \label{4.14}
\lim_{R\to\infty} I_2  =  \int_0^{\infty} {\rm d}x\, g_2(x,0)
\int_{-1}^1 \frac{{\rm d}t}{\sqrt{1-t^2}} = D_{\rm flat}
\end{equation}
Here, relation (\ref{A.2}) with $s_2=2\pi$ was applied at $x'=0$.
We conclude that
\begin{equation} \label{A.15}
\lim_{R\to\infty} D_{\rm disk} = 2 D_{\rm flat},
\quad \quad \nu = 2
\end{equation}

\subsection*{3D Sphere}
In 3D, the volume element ${\rm d}^3 r = r^2 {\rm d}r {\rm d}\Omega$,
where the angular part
${\rm d}\Omega = \sin\theta {\rm d}\theta {\rm d}\varphi$ with
$\theta \in (0,\pi)$ and $\varphi\in (0,2\pi)$.
Using the substitution $x=R-r$, the sphere dipole moment
(\ref{A.7}) reads
\begin{eqnarray}
D_{\rm sph} & = & \int_0^R {\rm d}x\, x (R-x)^2
\int {\rm d}\Omega\, \psi_{\rm sph}(x,\theta) \nonumber \\
& & + 2 \int_0^R {\rm d}x (R-x)^3 
\int {\rm d}\Omega\, \sin^2(\theta/2)
\psi_{\rm sph}(x,\theta) \label{A.16}
\end{eqnarray}
In the large-$R$ limit, the transformation (\ref{A.6}) implies
$D_{\rm sph} = I_1 + I_2$, where
\begin{eqnarray}
I_1 & = & \int_0^R {\rm d}x\, x\left( 1-\frac{x}{R} \right)^2
(2\pi) \int_0^{2R} {\rm d}y\, y \psi_{\rm flat}(x,y) \label{A.17} \\
I_2 & = & 2 \int_0^R {\rm d}x\, R\left( 1-\frac{x}{R} \right)^3
(2\pi) \int_0^{2R} {\rm d}y\, y \frac{y^2}{4 R^2} 
\psi_{\rm flat}(x,y) \label{A.18}
\end{eqnarray}
Here, we have used 
$\int {\rm d}\Omega = (2\pi/R^2)\int_0^{2R} {\rm d}y\, y$.
As in 2D, the $x/R$ terms are neglected as $R\to\infty$.
Thus, $\lim_{R\to\infty} I_1 = D_{\rm flat}$ and
\begin{equation} \label{A.19}
\lim_{R\to\infty} I_2 = 2\pi \int_0^{\infty} {\rm d}x
\lim_{R\to\infty} (2R)^3 \int_0^1 {\rm d}t\, t^3
\psi_{\rm flat}(x,2tR)
\end{equation} 
The leading asymptotic behavior
$\psi_{\rm flat}(x,2tR) \sim g_3(x,0)/(2tR)^3$ as $R\to\infty$
gives
\begin{equation} \label{A.20}
\lim_{R\to\infty} I_2 = 2\pi \int_0^{\infty} {\rm d}x\,
g_3(x,0) = D_{\rm flat}
\end{equation}
where the relation (\ref{A.2}) with $s_3=4\pi$ was applied
at $x'=0$.
Finally,
\begin{equation} \label{A.21}
\lim_{R\to\infty} D_{\rm sph} = 2 D_{\rm flat},
\quad \quad \nu=3
\end{equation}

\section*{ACKNOWLEDGMENTS}
The authors acknowledge support from the CNRS-SAS agreement,
Project No. 14439.
A partial support of L. \v Samaj by a VEGA grant is acknowledged.

\end{document}